\documentclass[12pt]{article}
\usepackage{graphicx}
\usepackage{epsf,epsfig,amsmath,amssymb,verbatim,mathrsfs}
\usepackage{epstopdf}
\def\beq{\begin{equation}}
\def\eeq{\end{equation}}
\def\bea{\begin{eqnarray}}
\def\eea{\end{eqnarray}}

\newcommand{\gsim}{\lower.7ex\hbox{$\;\stackrel{\textstyle>}{\sim}\;$}}
\newcommand{\lsim}{\lower.7ex\hbox{$\;\stackrel{\textstyle<}{\sim}\;$}}
\begin{document}
\pagestyle{empty}

\begin{center}
{\Large\bf Fermion Masses in Emergent Electroweak \\Symmetry Breaking}

\vspace{1cm}

{\sc Yanou Cui,$^{a,}$}\footnote{E-mail:  ycui@physics.harvard.edu}
{\sc Tony Gherghetta,$^{b,}$}\footnote{E-mail:  tgher@unimelb.edu.au}
{\sc\small and}
{\sc James Stokes$^{b,}$}\footnote{E-mail:  jstokes@physics.unimelb.edu.au}
\vspace{0.5cm}

{\it\small $^a$Jefferson Physical Laboratory, Harvard University,
Cambridge, Massachusetts 02138, USA\\
$^b$School of Physics, University of Melbourne, Victoria 3010, Australia}
\end{center}

\vspace{0.5cm}

\vspace{1cm}
\begin{abstract}
We consider the generation of fermion masses in an emergent model
of electroweak symmetry breaking with composite $W,Z$ gauge bosons.
A universal bulk fermion profile in a warped extra dimension is used for
all fermion flavors. Electroweak symmetry is broken at the UV (or Planck)
scale where boundary mass terms are added to generate the fermion flavor
structure. This leads to flavor-dependent nonuniversality in the gauge couplings.
The effects are suppressed for the light fermion generations but are enhanced
for the top quark where the $Zt{\bar t}$ and $Wt{\bar b}$ couplings can
deviate at the $10-20\%$ level in the minimal setup.
By the AdS/CFT correspondence our model implies that electroweak
symmetry is not a fundamental gauge symmetry. Instead the Standard Model with
massive fermions and $W,Z$ gauge bosons is an effective chiral Lagrangian for
some underlying confining strong dynamics at the TeV scale, where mass is generated
without a Higgs mechanism.

\end{abstract}

\newpage
\setcounter{page}{1}
\setcounter{footnote}{0}
\pagestyle{plain}

\section{Introduction}
The origin of mass of the elementary particles in the standard model remains a
long-standing problem. The simplest solution has been to invoke the Higgs mechanism
with the requisite Higgs boson. However the Higgs boson has not yet been discovered and
moreover, new physics at the TeV scale is required to stabilize the Higgs boson mass,
such as supersymmetry or strong dynamics. Therefore, even though the Higgs mechanism
seems essential it has yet to be experimentally verified. Recently, an alternative possibility
was considered, that does not rely on the Higgs mechanism~\cite{cgw}. Instead, analogous
to the $\rho$-mesons in QCD, the $W,Z$ bosons are composite, obtaining their mass from
the binding energy of some underlying strong dynamics. Electroweak symmetry is in fact
broken at the Planck scale, and the massive electroweak gauge bosons ``emerge" at low
energies when conformal symmetry is broken. Consequently, in the emergent model,
electroweak symmetry is not a fundamental gauge symmetry. Since the model is
phenomenological, parameters must be tuned to satisfy
electroweak precision tests, which may or may not depend on more fundamental physics.
Nonetheless, the emergent model shows that in the standard model the Higgs mechanism
is not mandatory. This idea is not new, with similar approaches having been previously
considered in~\cite{AGC}. However, it is distinct from other Higgsless ideas~\cite{tc, Csaki:2003zu},
where even though there is no Higgs boson, a Higgs mechanism is still present.

Besides the electroweak gauge bosons the fermions in the standard model
must also obtain a mass.  In Ref.~\cite{cgw}, the AdS/CFT correspondence was used
to construct a calculable model in a slice of AdS$_5$, where the fermions were treated as massless
and confined to the IR brane. This was a simplifying assumption to show that the correct
couplings to gauge bosons are indeed obtained. In this paper we show how fermion
masses are obtained in the emergent model with composite $W,Z$ bosons. As outlined
in Ref.~\cite{cgw} the fermions must have a universal bulk profile in order to guarantee
gauge coupling universality. Since electroweak symmetry is broken on the UV brane,
boundary mass terms will then generate the necessary flavor structure. The fermion bulk
profiles do obtain flavor-dependent corrections from non-universal UV boundary masses, but since both the
electroweak gauge bosons and fermions are peaked near the IR brane, the nonuniversality
in the gauge couplings is suppressed. These effects are greatest for the fermions of the
third generation, where in particular for the top quark, deviations in gauge coupling universality
are at the $10-20\%$ level. While there are no conflicts with current direct bounds, this
deviation in the top-quark coupling causes a tension with indirect
bounds from electroweak precision data. One way to alleviate this tension is to treat the top
quark separately from the light fermions as has been considered in other Higgsless models.

Of course, the fermion mass hierarchy will no longer arise from the wavefunction overlap.
Instead one must rely on a Froggatt-Nielsen mechanism~\cite{fn} on the UV brane to generate the
Yukawa coupling hierarchy. Nevertheless the extra dimension still plays a role in reducing
the possible hierarchy in Yukawa couplings.
A UV localized fermion profile with a large coupling to the UV brane would require much smaller
Yukawa couplings compared to an IR localized fermion profile that is naturally suppressed near the
UV brane. Furthermore, the Kaluza-Klein (KK) states are sufficiently heavy but also more suppressed at the
IR brane due to large IR boundary kinetic terms. Interestingly, the lightest KK state of the
top quark is approximately 1.5 TeV. With this sufficiently small mass and sizeable couplings it
can be searched for at the LHC.

The outline of this paper is as follows. In Section 2 we review the emergent model of Ref.~\cite{cgw}
and introduce bulk fermions with a universal profile. The matching to Standard Model (SM) gauge
couplings is discussed in Section 3. This includes the photon, $W$-boson and $Z$-boson couplings.
An electroweak precision analysis is also presented. The conclusion and final comments are given
in Section 4.

\section{The 5D Emergent Model}
\label{sec5dmodel}
The emergent model of electroweak symmetry breaking is defined in five
dimensions using the Randall-Sundrum framework~\cite{Randall:1999ee}.
Consider a slice of AdS$_5$ with five-dimensional (5D) metric
\beq
ds^2=\frac{1}{(kz)^2}(\eta_{\mu\nu}dx^\mu dx^\nu+dz^2)\equiv g_{MN}dx^Mdx^N~,
\eeq
where $k$ is the AdS curvature scale. The 5D spacetime indices are written
as $M=(\mu,5)$, with $\mu=0,1,2,3$, and $\eta_{\mu\nu}={\rm diag}(-+++)$
is the four-dimensional (4D) Minkowski metric. The fifth dimension $z$ is compactified on a $Z_2$
orbifold, with a UV (IR) brane located at the fixed point $z^*=z_{UV}(z_{IR})$.
The $z$ coordinate is related to the 4D energy scale, and
the scale of the UV (IR) brane is chosen to be $z_{UV}=k^{-1},
(z_{IR}={\cal O}({\rm TeV}^{-1}))$ respectively, where
$k\simeq M_P=2.4\times 10^{18}$ GeV is the reduced Planck scale.
An underlying stabilization mechanism (such as in Ref.~\cite{gw}) is assumed
for the separation between the two branes.

The 5D emergent model~\cite{cgw} contains an electroweak symmetry,
$SU(2)_L\times U(1)_Y$, which is preserved in the 5D bulk and on the IR
brane but broken down to the electromagnetic gauge group $U(1)_Q$ on
the UV brane. In addition brane kinetic terms~\cite{BKT} are added to the boundaries to
ensure that the $W,Z$ gauge bosons are identified with the lowest-lying KK
modes which are peaked towards the IR brane and are sufficiently separated
from the higher KK modes. Since the breaking on the UV brane is implemented
via Dirichlet boundary conditions, the model is effectively Higgsless at low energies.
It differs from the usual Higgsless model~\cite{Csaki:2003zu} in that the $W,Z$
bosons are identified as the lowest-lying KK states peaking towards the IR brane (as opposed
to the flat zero modes) and therefore via the AdS/CFT dictionary~\cite{Maldacena:1997re,
adscftdict} are composite. On the other hand the massless gauge bosons such as the
gluons are elementary fields as well as the photon which is (mostly) elementary.

\subsection{Fermions in the composite $W/Z$ model}
In the original model the fermions were simply treated as confined to the IR brane.
To obtain a more realistic scenario we will suppose that the fermions are bulk fields
which obtain their mass from electroweak symmetry breaking on the UV brane.
However, since the profiles of the $W,Z$ gauge bosons are no longer constant
(being localized towards the IR brane), gauge coupling universality is no longer automatic.
Instead to guarantee universality, the fermion bulk profiles must be identical for all flavors.
On the UV boundary fermion masses are then introduced to generate the required
flavor structure. A large modification of the fermion profiles occurs near the UV brane, but
since the $W,Z$ gauge boson profiles are suppressed there, the deviations from gauge coupling
universality of the light fermions are naturally tiny and are compatible with current experimental
constraints.

For each SM fermion flavor $i$, we introduce two Dirac fermions $\Psi_i^{(L)},\Psi_i^{(R)}$ where
$L (R)$ refer to the left (right)-handed $SU(2)_L$ doublets (singlets) in the Standard Model.
The fermion action is given by
\begin{equation}
S_\Psi = i\int d^5 x \sqrt{-g} \left[\frac{1}{2}(\overline{\Psi}_i^{(L)} \Gamma^M D_M \Psi_i^{(L)} -
D_M {\overline\Psi}_i^{(L)}\Gamma^M \Psi_i^{\rm (L)}) + m_L^{(i)} \overline{\Psi}_i^{(L)}\Psi_i^{(L)}
+ (L\leftrightarrow R) \right].
\end{equation}
The curved-space gamma matrices $\Gamma^M = e^M_A \gamma^A$, where $e^M_A = (kz)
\delta^M_A$ is the vielbein and $\gamma^A$ are the gamma matrices in flat space given by
\begin{eqnarray}
\gamma^\mu = -i \left(
\begin{array}{cc}
0 & \bar\sigma^\mu \\
\sigma^\mu & 0
\end{array}\right),
& &
\gamma^5 =  \left(
\begin{array}{cc}
1 & 0 \\
0 & -1
\end{array}
\right),
\end{eqnarray}
with $\sigma^\mu (\bar\sigma^\mu) = (1, \pm\sigma^i)$ and $\sigma^i$ are the usual Pauli matrices.
The spinor covariant
derivative $D_M= \partial_M+\omega_M$, where $\omega_M= (-\frac{1}{2z}\gamma_\mu\gamma^5,0)$
is the spin connection. The bulk fermion masses $m_{L,R}^{(i)} = c_{L,R}^{(i)}k$ are parametrised in
units of the curvature scale, $k$ with dimensionless parameters $c_{L,R}^{(i)}$. To maintain gauge coupling universality in the bulk, we require the left and right-handed fermion wavefunctions to be the same and universal for all fermion flavors, so we define $c_R^{(i)} = -c_L^{(i)} \equiv c$.
In addition the Dirac fermions $\Psi_i$ (suppressing $L, R$ labels) are
decomposed into their left$(-)$ and right$(+)$-handed Weyl components by writing  $\Psi_i= \Psi_{i+}+
\Psi_{i-}$, where $\Psi_{i\pm} =\pm \gamma_5 \Psi_{i\pm}$. The (massless) zero modes are then taken
to be associated with  $\Psi^{(L)}_{i-}$ and $\Psi^{(R)}_{i+}$.

To generate masses for the 4D fermion zero modes we will consider adding a UV boundary
mass term, with an action of the form
\begin{equation}
        S_m^{(UV)} = i\int d^5 x \sqrt{-g} \, \lambda_5^{(i)}  \left[{\overline\Psi}_i^{(L)} \Psi_i^{(R)}
        +{\overline\Psi}_i^{(R)} \Psi_i^{(L)}\right] (kz)\delta(z - z_{\rm UV}),
\label{massLag}
\end{equation}
where $\lambda^{(i)}_5$ are flavor-dependent Yukawa couplings. This mass term will introduce a
flavor-dependent modification to the fermion bulk profile which for light fermions is proportional to
$(m_i/m_{KK})^2$.
However, note that the fermion mass hierarchy now arises from the boundary Yukawa couplings and is no
longer explained by the wavefunction overlap in the bulk. Instead the hierarchy in $\lambda_5^{(i)}$ is
assumed to arise from a Froggatt-Nielsen mechanism~\cite{fn} on the UV brane.

Furthermore, fermion kinetic terms can be added on the IR brane. The boundary action is given by
\begin{eqnarray}
S_{KE}^{(IR)} &=& i\int d^5 x \sqrt{-g} \left[\frac{1}{2}\eta_{iL} (\overline{\Psi}_i^{(L)} \Gamma^\mu D_\mu \Psi_i^{(L)} -
D_\mu {\overline\Psi}_i^{(L)}\Gamma^\mu \Psi_i^{\rm (L)}) + (L\leftrightarrow R) \right] (kz)\delta(z-z_{IR}),\nonumber\\
&=&i\int d^5 x \sqrt{-g} \left[\eta_{iL} \overline{\Psi}_i^{(L)} e^\mu_\alpha \gamma^\alpha \partial_\mu \Psi_i^{(L)} +
(L\leftrightarrow R) \right] (kz)\delta(z-z_{IR}),\label{IRKE}
\end{eqnarray}
where $\eta_{iL,R}$ are constant coefficients of the brane kinetic terms. Since the fermion bulk profiles
are universal these coefficients are also assumed to be flavor-independent with $\eta_{iL}=\eta_{iR}\equiv\eta$.
In addition, we will see later that boundary kinetic terms help to achieve 5D perturbativity.
Note also that boundary kinetic terms can be added on the UV boundary but these have a negligible effect,
and for simplicity we will neglect them.

The equations of motion and boundary conditions can be obtained by varying the total action,
$S_\Psi+S_{m}^{(UV)}+S_{KE}^{(IR)}$. They are solved by using the Kaluza-Klein decomposition
\begin{equation}
	\Psi_\pm^{(L),(R)} (x^\mu,z)  = \sum_{n=0}^{\infty} f_{L\pm,R\pm}^{(n)}(z) \psi_\pm^{(n)}(x^\mu),
	\label{kksum}
\end{equation}
where we have suppressed the flavor index $i$, and assumed identical 4D modes $\psi_\pm^{(n)}$
for $\Psi^{(L),(R)}$. The form of the profile functions $f_\pm^{(n)}$ (suppressing $L,R$ labels) are found to be
\begin{equation}
	f_\pm^{(n)}(z)  = N_n (kz)^{5/2} [J_{\alpha_\pm}(m_n z) + b_n Y_{\alpha_\pm}(m_n z) ],
	\label{fermprofile}
\end{equation}
where $\alpha_\pm \equiv c\pm 1/2$ and $N_n, b_n$ are the same constants for both $\pm$ . The
vanishing of the boundary variation leads to the following set of boundary conditions
\begin{align}
	f^{(n)}_{L+}(z_{\rm UV}^+) = -\lambda_5 f^{(n)}_{R+}(z_{\rm UV}), \quad f^{(n)}_{L+}(z_{\rm IR}^-)
	= -(\eta_L k)(m_n z_{IR})f_{L-}^{(n)}(z_{IR}),\label{newbc1} \\
	f^{(n)}_{R-}(z_{\rm UV}^+) = \lambda_5 f^{(n)}_{L-}(z_{\rm UV}), \quad f^{(n)}_{R-}(z_{\rm IR}^-) =
	(\eta_R k)(m_n z_{IR})f_{R+}^{(n)}(z_{IR}),
	\label{newbc2}
\end{align}
where $\delta\Psi_+^{(L)}=\delta\Psi_-^{(R)}=0$ on the boundaries. The profiles $f_{L+, R-}^{(n)}$ are
discontinuous at the boundaries and we define $f(z_{\rm UV}^+)=\lim_{\epsilon\rightarrow 0}
f(z_{\rm UV}+\epsilon)$ and $f(z_{IR}^-)=\lim_{\epsilon\rightarrow 0} f(z_{\rm IR}-\epsilon)$. This definition
is necessary because boundary terms arising from the integration by parts of bulk kinetic terms assumes a continuous function limit. Note that we have used $\int_{z_{UV}}^{z_{IR}} dz \,\delta(z-z_{UV}) f(z) = \frac{1}{2}f(z_{UV})$ and $\int_{z_{UV}}^{z_{IR}} dz \,\delta(z-z_{IR}) f(z) = \frac{1}{2}f(z_{IR})$.

The fermion profiles (\ref{fermprofile}) depend on three parameters $N_n, b_n$ and $m_n$. These are determined
by the two boundary conditions and a normalization condition. First, imposing the IR boundary condition determines the constants $b_n^{(L),(R)}$ to be
\begin{eqnarray}
	b_n^{(L)} &=& -\frac{J_{\alpha_{L+}}({\widehat m}_n) +(\eta_L k)\, {\widehat m}_n J_{\alpha_{L-}}({\widehat m}_n)}{Y_{\alpha_{L+}}({\widehat m}_n)+ (\eta_L k)\, {\widehat m}_n Y_{\alpha_{L-}}({\widehat m}_n)},\\
	b_n^{(R)} &=& -\frac{J_{\alpha_{R-}}({\widehat m}_n) -(\eta_R k)\, {\widehat m}_n J_{\alpha_{R+}}({\widehat m}_n)}{Y_{\alpha_{R-}}({\widehat m}_n)-(\eta_R k)\, {\widehat m}_n Y_{\alpha_{R+}}({\widehat m}_n)},
\end{eqnarray}
where ${\widehat m}_n = m_n z_{\rm IR}$. The mass eigenvalues, $m_n$ are then obtained by imposing the UV boundary conditions. This leads to the algebraic equation
\begin{eqnarray}
      &&(\lambda_5)^2 \left[J_{\alpha_{R+}}(m_n z_{\rm UV}) + b_n^{(R)} Y_{\alpha_{R+}}(m_n z_{\rm UV})\right]
      \left[J_{\alpha_{L-}}(m_n z_{\rm UV}) + b_n^{(L)} Y_{\alpha_{L-}}(m_n z_{\rm UV})\right] \nonumber\\
      &&\qquad\quad=\left[J_{\alpha_{R-}}(m_n z_{\rm UV}) + b_n^{(R)} Y_{\alpha_{R-}}(m_n z_{\rm UV})\right]
      \left[J_{\alpha_{L+}}(m_n z_{\rm UV}) + b_n^{(L)} Y_{\alpha_{L+}}(m_n z_{\rm UV})\right],
      \label{masseqn}
\end{eqnarray}
which is solved numerically.

In order to obtain the correct dimensionally reduced theory from the KK expansion of the action, namely
\begin{equation}\label{e:dimred}
	S_4 = i\int d^4 x \, (\overline{\psi}^{(n)} \gamma^\mu \partial_\mu \psi^{(n)} + m_n \overline{\psi}^{(n)}
	\psi^{(n)}),
\end{equation}
with $\psi^{(n)}=\psi_+^{(n)}+\psi_-^{(n)}$,
one must impose suitable normalization conditions on the profile functions. The fermion boundary
conditions \eqref{newbc1} and \eqref{newbc2} for the Sturm-Liouville problem leads to the more general
normalization condition (suppressing $L,R$ labels)
\begin{equation}
     \int \frac{dz}{(kz)^4}\, f_\pm^{(m)}(z) f_\pm^{(n)}(z) + \frac{\eta}{2(kz_{IR})^3}f_{\rm even}^{(m)}(z_{IR})
     f_{\rm even}^{(n)}(z_{IR})= \frac{1}{2}\delta_{mn} + \Delta_{mn}^\pm,
     \label{e:normgen}
\end{equation}
where the $\eta$ term only exists for the even mode.
Compared to the usual normalization condition, Sturm-Liouville orthonormality requires two additional
pieces in (\ref{e:normgen}): the $\eta$ term on the left-hand side due to the presence of boundary kinetic terms, and $\Delta_{mn}^\pm$ on the right-hand side due to boundary mass terms. The necessity of the
$\Delta_{mn}^\pm$ term, due to boundary mass terms, was first pointed out in \cite{neubert}.
Note that analogous to the flavor factor $a_n,b_n$ in \cite{neubert}, the  $\delta_{mn}$ term in (\ref{e:normgen}) includes an additional factor $\frac{1}{2}$, to ensure a canonical kinetic term for the KK modes. Using our boundary conditions and, for simplicity, ignoring off-diagonal entries (flavor mixings) in the Yukawa matrices (as considered in \cite{neubert}), we obtain for $m\neq n$
\bea
\label{delLm}
\Delta_{mn}^{L-}&=&\frac{m_n}{m_m^2-m_n^2}\left[\frac{1}{(kz_{UV})^4}f^{(m)}_{L-}(z_{UV})f^{(n)}_{L+}(z_{UV})-\frac{1}{4(kz_{IR})^4}\left(3+\frac{m_m^2}{m_n^2}\right)f^{(m)}_{L-}(z_{IR})f^{(n)}_{L+}(z_{IR})\right]\nonumber\\
&&\qquad\qquad\qquad+\left(m\leftrightarrow n\right)~,\\
\Delta_{mn}^{L+}&=&\frac{-m_n}{m_m^2-m_n^2}\left[\frac{1}{(kz_{UV})^4}f^{(m)}_{L+}(z_{UV})f^{(n)}_{L-}(z_{UV})-\frac{1}{(kz_{IR})^4}f^{(m)}_{L+}(z_{IR})f^{(n)}_{L-}(z_{IR})\right]+\left(m\leftrightarrow n\right).\nonumber\\
\eea
In (\ref{delLm}) we have used the boundary condition (\ref{newbc1}) to rewrite the $\eta$ term.
Similar expressions are obtained for the $(R)$ fields, except for an overall sign difference.
When $m=n$, the expressions are given by
\beq
    \Delta_{nn}^{L-,R-}=-\Delta_{nn}^{L+,R+}=\frac{1}{2m_n}\left[\frac{1}{2(kz_{IR})^4}f_{-}^{(n)}(z_{IR})
    f_{+}^{(n)}(z_{IR})-\frac{1}{(kz_{UV})^4}f_{-}^{(n)}(z_{UV})f_{+}^{(n)}(z_{UV})\right].
\eeq
Given that $\Delta_{nn}^+=-\Delta_{nn}^-$, the normalization factors, $N_n$ are most simply determined by
summing together the $\pm$ expressions in (\ref{e:normgen}) to give
\begin{equation}
      \int \frac{dz}{(kz)^4} \left( f_{L-}^{(n)}(z) f_{L-}^{(n)}(z)+f_{L+}^{(n)}(z) f_{L+}^{(n)}(z)\right)
      +\frac{\eta}{2(kz_{IR})^3}f_{L-}^{(n)}(z_{IR}) f_{L-}^{(n)}(z_{IR})=1\,,
\label{normfactor}
\end{equation}
where for the $(R)$ fields the boundary kinetic term part contains $f_{R+}^{(n)}$. This relation is also
consistent with having canonically normalized kinetic terms in the 4D theory.

\subsection{Example: Massless bulk fermions $(c=0)$}
\label{c0example}
An interesting case to consider for the bulk fermion masses is when $c=0$. Besides obtaining
simple analytic expressions for the fermion profiles, having $c$ near $0$ naturally accommodates
a TeV scale mass with an ${\cal O}(1)$ UV Yukawa coupling. The fermion mass hierarchy is then obtained
by UV Yukawa couplings which are similar to those in the SM. Assuming $\eta_L=\eta_R\equiv \eta$ the
fermion profiles are given by
\begin{eqnarray}
\label{f0even}
          |f_{L-}^{(n)}(z)| = |f_{R+}^{(n)}(z)| &=& N_n^{(0)}\,(kz)^2 \left[\cos({\widehat m}_n- m_n z) -
    (\eta k){\widehat m}_n\sin({\widehat m}_n- m_n z)\right],\\
           |f_{L+}^{(n)}(z)| = |f_{R-}^{(n)}(z)| &=& N_n^{(0)}\,(kz)^2 \left[\sin({\widehat m}_n-m_n z) +
    (\eta k){\widehat m}_n\cos({\widehat m}_n- m_n z)\right],
    \label{f0odd}
\end{eqnarray}
where $N_n^{(0)}$ is the normalization constant. Using (\ref{normfactor}) it  is given by
\begin{equation}
      N_n^{(0)} \simeq \frac{1}{\sqrt{z_{\rm IR}}}\sqrt{\frac{1}{1+ (\eta k)/2+(\eta k)^2 \widehat{m}_n^2 }}~,
      \label{fermionnorm}
\end{equation}
where $z_{\rm UV}/z_{\rm IR} \ll 1$.
It is a useful check to consider the limit $m_0\rightarrow 0$ corresponding to turning off the boundary
Dirac mass $(\lambda_5=0)$. In this case we find that  $| f_{L-}^{(0)}(z)| = |f_{R+}^{(0)}(z)|\propto (kz)^2 $
which is just the even zero-mode fermion wavefunction, while $| f_{L+}^{(0)}(z)| = |f_{R-}^{(0)}(z)| \rightarrow 0$,
represents the odd zero-mode fermion that is projected out by the boundary conditions.

The mass eigenvalues are obtained by solving the equation (\ref{masseqn}), which in the limit $\pi k R \gg  1$,
simply becomes
\begin{equation}
      \lambda_5^2 = \left(\frac{\tan {\widehat m}_n + (\eta k) {\widehat m}_n}{1-(\eta k) {\widehat m}_n
      \tan {\widehat m}_n}\right)^2.
\end{equation}
In the limit $\lambda_5\ll 1$ the smallest solution to this equation is approximately given by
\begin{equation}
	m_{i0} \simeq \frac{\lambda_5^{(i)}}{\sqrt{(1+\eta k)^2+2\eta k(\lambda_5^{(i)})^2}} z_{\rm IR}^{-1},
	\label{newm0}
\end{equation}
where the flavor dependence has been reintroduced. Assuming $\lambda_5^{(i)}\ll 1$, the next-heaviest KK mode is near ${\widehat m}_1\sim \pi/2$. However, when $\lambda_5^{(i)}={\cal O}(1)$ the first KK mode becomes light $\widehat m_1 <1$ and is approximately given by
\begin{equation}
        m_{i1} \simeq \sqrt{  \frac{1}{(\lambda_5^{(i)})^2}+\frac{2}{(\eta k)}} z_{\rm IR}^{-1}.
\end{equation}
This can cause the next-heaviest KK mode of the top quark to be too light and violate experimental bounds. For $z_{\rm IR}^{-1}=1800$ GeV, $\lambda_5^{(t)}=1.15$ and $\eta k=10$  we numerically obtain $m_{t0}\simeq 171$ GeV and $m_{t1}\simeq 1503$ GeV. This compares with the electron where $\lambda_5^{(e)}=3.1\times 10^{-6}$,
numerically gives $m_{e0}\simeq 0.5$ MeV and $m_{e1}\simeq 2938$ GeV. The remaining light fermion masses
are obtained by choosing $\lambda_5^{(i)}\simeq  (m_i/m_e) \lambda_5^{(e)}$. The intriguing feature of the
mass spectrum is that unlike most existing RS models, the first KK top quark in our model can be as light
as 1.5 TeV, with a sizeable coupling to the SM top quark:
$g_{t^{(0)}t^{(1)}Z}\sim \frac{1}{3} g_{t^{(0)}t^{(0)}Z}$.
The KK top quarks are pair-produced and give rise to a $2t+2Z$ signal.
The analysis of this signal will be left for future work, but could be an interesting discovery channel for our model.

\section{Matching to the SM gauge couplings}
The gauge couplings are obtained from the wavefunction overlap between the fermion profiles and the
composite $W,Z$-boson profiles. While most of the bulk overlap will be universal, the gauge couplings will receive small nonuniversal contributions from the UV Yukawa coupling. This arises from
the nonuniversal deformation of the fermion profile near the IR brane, where the fields are mostly localized.
Assuming a gauge-covariant derivative $D_M=\partial_M+i g_{5L} A_M^a T^a + i g_{5Y} Y_5 B_\mu$,  and working in the gauge where $A_5 \equiv 0$, the bulk interaction term is given by
\begin{eqnarray}
     S_{int} &=& i\int d^5 x \sqrt{-g} \, (kz) \left[ \frac{1}{\sqrt{2}}g_{5L} \overline{\Psi}^{\rm (L)} \gamma^\mu
     (A_\mu^{\rm L+} T^+ + A_\mu^{\rm L-} T^-)\Psi^{\rm (L)} + g_{5L} \overline{\Psi}^{\rm (L)} \gamma^\mu
     A_\mu^{L3} T^3 \Psi^{\rm (L)}\right.\nonumber\\
     \qquad &&\left. + g_{5Y} \overline{\Psi}^{\rm (L)} \gamma^\mu B_\mu Y_{5L} \Psi^{\rm (L)} + g_{5Y}
                    \overline{\Psi}^{\rm (R)} \gamma^\mu B_\mu Y_{5R} \Psi^{\rm (R)} \right],
        \label{gaugeLag}
\end{eqnarray}
where $T^\pm,T^3$ are the SU(2)$_L$ generators, $Y_{5L,R}$ is the 5D hypercharge
and  $A_\mu^{L\pm}=1/\sqrt{2}(A_\mu^{L1}\mp i A_\mu^{L2})$.

In addition to the bulk interaction terms we also need to include boundary interaction terms. These terms
are in fact \textit{required} by the renormalization principle in quantum field theory for models with brane
kinetic terms. Their value can be fixed by the brane kinetic term coefficients and the 5D cutoff scale. The inclusion of these terms is also a more systematic and accurate interpretation of the `rescaling factor' in Eqs.(36),(37) of \cite{cgw}, which is crucial to ensure 5D perturbativity in models with large brane kinetic terms. An important starting point is to note that brane kinetic terms are essentially field renormalization on the brane~\cite{Dvali:2000rx, Georgi:2000ks}. In a consistent renormalized theory, whenever a field gets renormalized, the same scaling factor should enter all interaction terms involving that field.

For concreteness, let us first briefly review QED as a renormalized perturbation theory. We start with the Lagrangian written in terms of the bare coupling $(g_0)$ and fields $(A_\mu, \Psi)$
\beq
\mathcal{L}=\bar{\Psi}\gamma^{\mu}(\partial_\mu+ i g_0 A_{\mu})\Psi+\frac{1}{4}(\partial_\mu A_{\nu}-\partial_\nu A_{\mu})^2.
\label{bareLag}
\eeq
Upon renormalization, all bare quantities are rewritten in terms of the `physical' or `renormalized' quantites
(labeled with the superscript $(r)$)
\beq
\mathcal{L}=Z_g i g^{(r)}\bar{\Psi}^{(r)}\gamma^{\mu}A^{(r)}_{\mu}\Psi^{(r)}+Z_{\Psi}\bar{\Psi}^{(r)}\gamma^{\mu}\partial_\mu\Psi^{(r)}+Z_A\frac{1}{4}(\partial_\mu A^{(r)}_{\nu}-\partial_\nu A^{(r)}_{\mu})^2,
\label{renormLag}
\eeq
with renormalization factors $Z_g, Z_\Psi, Z_A$. Comparing the two expressions (\ref{bareLag}) and
(\ref{renormLag}) the bare and renormalized quantities are related by:
$A^{(r)}_\mu=Z_A^{-1/2}A_\mu,\Psi^{(r)}=Z_{\Psi}^{-1/2}\Psi, g^{(r)}=Z_A^{1/2}Z_\Psi Z_g^{-1}g_0$. Furthermore, gauge invariance requires the two parts of the covariant derivative to be renormalized by the same factor, i.e. $Z_\Psi=Z_g$, so that $g^{(r)}=Z_A^{1/2}g_0$.

Now let us consider the brane kinetic and interaction terms in the 5D model. The bare coupling and
fields in the QED example are analogous to the bulk coupling and fields. The presence of brane kinetic terms
implies that brane dynamics renormalizes the fields and coupling. In particular, analogous to the relation
$Z_g g^{(r)}=Z_A^{1/2}Z_\Psi g_0$, the `physical' brane coupling is not only related to the original 5D coupling
(analogous to $g_0$), but also contains large renormalization factors. The brane kinetic terms in our 5D model lead to an effective field renormalization for the gauge field which is formally written as\footnote{The extra factor of $2$ here is chosen to be consistent with the boundary kinetic term convention used for the gauge field in Ref.~\cite{cgw}, where $\zeta_A$ absorbs a factor of $1/2$ from the $\delta$-function integration compared to the $\eta$ convention defined in Eq.(\ref{IRKE})} $Z_A=1+2\zeta_A\,\delta(z-z_i)$ where $i=\rm IR, UV$. Similarly for the boundary fermion kinetic term where
$Z_f=1+\eta\,\delta(z-z_{IR})$. With a canonically normalized Kaluza-Klein fermion, the coefficient
$\xi_{IR}$ of the IR boundary interaction term must therefore be matched as\footnote{We have included
a factor of $2$ in front of $\xi_{IR}$ to cancel the $\frac{1}{2}$ factor coming from the $\delta$-function
integration.}: $\int dz\,2\,\xi_{IR}\delta(z-z_{IR})=\int dz\,(Z_f Z_A^{1/2}-1)=\int dz\,\left[(1+\eta\,\delta(z-z_{IR}))
\sqrt{1+2\zeta_A\delta(z-z_{IR})}-1\right]$.
Since we evaluate the above expression at $z=z_{IR}$, the substitution $\delta(z-z_{IR})\rightarrow\delta(0)$ can be made. According to the Fourier transform of the $\delta$-function, we can replace $\delta(0)$ with $\Lambda$- the 5D momentum cutoff, which eventually leads to the relation
\begin{equation}
   \xi_{IR}=\frac{1}{2\Lambda}\left[(1+\eta\Lambda)\sqrt{1+2\zeta_L\Lambda}-1\right].
   \label{xiIReqn}
\end{equation}
Note that in the limit $\eta\Lambda\gg 1$ and $\zeta_L\Lambda\gg 1$ the expression simply becomes
$\xi_{IR}\simeq\left(\frac{\zeta_L}{2}\Lambda\right)^{1/2}\eta$.

It is important to note that on the IR brane, $U(1)_Y$ and $SU(2)_L$ can have different brane kinetic term coefficients, so the induced brane interactions are proportional to $\sqrt{\zeta_Y}, \sqrt{\zeta_L}$ respectively. However it is easy to check that in order to ensure a universal 4D electric charge $Q$ for fields with the same
$T^3$ and $Y$, the renormalization factors for the $U(1)_Y$ and $SU(2)_L$ brane interactions should be the same. This requires scaling the hypercharge $Y$ on the IR brane by an amount $\sqrt{\zeta_L/\zeta_Y}$. Using the $\zeta_L,\zeta_Y$ values in \cite{cgw}, which is the benchmark point for a good fit to electroweak precision observables, we find that the rescaling is about $70$. As we will see, the SM gauge couplings are matched with a bulk coupling $g_{5Y}\sqrt{k}\sim 0.1$. This is sufficiently small that even the IR rescaled coupling remains perturbative.

Now we are ready to write down the boundary interaction terms with coefficients $\xi_{IR},\xi_{UV}$
\begin{eqnarray}
      S_{int}^{(4D)} &=& i\int d^5 x \sqrt{-g} \, \bigg\{2\, \xi_{UV}\,\delta(z-z_{UV}) (kz)^2  e_5 Q_5
      \left[\overline{\Psi}^{\rm (L)} \gamma^\mu A_\mu^{(5)} \Psi^{\rm (L)} +
    \overline{\Psi}^{\rm (R)} \gamma^\mu A_\mu^{(5)}  \Psi^{\rm (R)} \right]\nonumber\\
&&\qquad\qquad\qquad+ 2\,\xi_{IR}\,\delta(z-z_{IR}) (kz)^2 \left[ \frac{1}{\sqrt{2}}g_{5L} \overline{\Psi}^{\rm (L)} \gamma^\mu
(A_\mu^{\rm L+} T^+ + A_\mu^{\rm L-} T^-)\Psi^{\rm (L)}\right.\nonumber\\
  && \quad \left. +\, g_{5L} \overline{\Psi}^{\rm (L)} \gamma^\mu A_\mu^{L3} T^3 \Psi^{\rm (L)} + g_{5Y} \overline{\Psi}^{\rm (L)} \gamma^\mu B_\mu Y_{5L} \Psi^{\rm (L)} + g_{5Y}
  \overline{\Psi}^{\rm (R)} \gamma^\mu B_\mu Y_{5R} \Psi^{\rm (R)} \right]\bigg\},
\label{boundaryintLag}
\end{eqnarray}
where $Q_5$ is the 5D electric charge and
\begin{equation}
   A_\mu^{(5)} =  \frac{1}{\sqrt{g_{5L}^2+g_{5Y}^2}}(g_{5Y} A_\mu^{L3} + g_{5L} B_\mu);      \qquad
   e_5=\frac{g_{5L} g_{5Y}}{\sqrt{g_{5L}^2+g_{5Y}^2}}.
\end{equation}
Note that in (\ref{boundaryintLag}) the interaction has been written to show the dependence on the ``bare" 5D coupling, $g_5$. The effective 4D gauge couplings are obtained by substituting the Kaluza-Klein expansions in the total action $S_{int} + S_{int}^{(4D)}$.  The KK decomposition of the fermions is given in (\ref{kksum}) while that of the gauge fields is \cite{cgw}
\begin{eqnarray}
A_\mu^{L3}(x,z)&=&f_{L3}^{(0)}(z) A_\mu(x)+\displaystyle\sum_{n=1}^{\infty}f_{L3}^{(n)}(z)Z_\mu^{(n)}(x),
\label{KKdec1}\\
B_\mu(x,z)&=&f_B^{(0)}(z) A_\mu(x)+\displaystyle\sum_{n=1}^{\infty}f_B^{(n)}(z)Z_\mu^{(n)}(x),
\label{KKdec2}\\
A_\mu^{L\pm}(x,z)&=&\displaystyle\sum_{n=1}^{\infty}f_W^{(n)}(z)W_\mu^{(n)\pm}(x),
\label{KKdec3}
\end{eqnarray}
where $A_\mu(x)$ is the 4D photon. However, note that the fermion profiles $f_{L+}^{(0)}$ and $f_{R-}^{(0)}$ are discontinuous at the boundaries and are assumed to vanish at $z_{\rm UV}$ and  $z_{\rm IR}$. On a $Z_2$ orbifold they would correspond to odd fields. Therefore the boundary contributions to the gauge couplings only arise from $f_{L-}^{(0)}$ and $f_{R+}^{(0)}$ (which correspond to even fields on a $Z_2$ orbifold).

It is instructive to consider the limit $\eta,\zeta\rightarrow \infty$. As shown in Ref.\cite{cgw},
the $\zeta\rightarrow\infty$ limit causes massless $W,Z$ bosons with $f_A\propto \frac{1}{\sqrt{\zeta_A}}$
to be completely localized on the IR brane, while the higher KK modes decouple. This is also true for the bulk fermions considered in Section 2.2. Using (\ref{fermionnorm}) the normalization integral (\ref{normfactor}) for fermion KK modes is dominated  by the bulk integral. The massive KK modes are therefore predominantly
located in the bulk. However for a massless fermion mode with $f_\psi \propto \frac{1}{\sqrt{\eta}}$, the normalization integral (\ref{normfactor}) is dominated by the IR boundary term. On the
IR brane an order one coupling then arises from the overlap integral
$\propto \sqrt{\zeta_A} \eta  f_\psi f_\psi f_A$. This realizes the original Randall-Sundrum limit~\cite{Randall:1999ee} where massless gauge bosons interact weakly with massless fermions on the IR brane.

In the emergent electroweak model the parameters ($\eta$, $\zeta_L$, $\Lambda$) must be chosen to match the SM couplings at the level required by precision electroweak data. In Ref.~\cite{cgw} it was shown that with the simplified assumption of all fermions confined to the IR brane, the emergent model does match the SM well
and the $S,T$ parameters are compatible with the LEP bounds. This is due to the fact that there is a
built-in custodial symmetry, together with a sufficient separation between the $W,Z$-boson masses and the IR mass scale, as well as suppressed KK
couplings due to the presence of large brane kinetic terms. To again be compatible with electroweak precision tests, the parameters of the more realistic model with fermion masses presented here, should be chosen to mimic the simplified model as close as possible. Furthermore, the gauge coupling to fermions should match well with the self-interaction couplings of non-abelian gauge fields, which is constrained by LEP data to only allow deviations at the few percent level~\cite{Alcaraz:2006mx}. As we will see, all these constraints can be satisfied by minimizing the bulk contribution to the fermion-gauge boson couplings. This is straightforward to achieve by using $\eta k\sim{\cal{O}}(100)$ for all the light fermions, while the top quark is an exception due to its large mass. Note that
large brane kinetic terms are required, but as shown in Ref.~\cite{Ponton:2001hq} they are perturbatively consistent. We will first study the matching to the photon coupling and fit numerical input parameters based on this most constraining consideration. Then we will discuss the matching to the $W, Z$-boson couplings.

\subsection{Photon coupling}
In the emergent model the photon profiles are $f_{L3}^{(0)}=N_A^{(0)}/g_{5L}, f_B^{(0)}=N_A^{(0)}/g_{5Y}$
where $N_A^{(0)}$ is a normalization constant (see Eq.(33) in Ref.~\cite{cgw}). Therefore, the coupling of the 4D photon $A_\mu(x)$ to the fermion
KK modes is obtained by substituting \eqref{KKdec1} and \eqref{KKdec2} into (\ref{gaugeLag}) and
(\ref{boundaryintLag}). The bulk contribution becomes
\begin{eqnarray}
 S_{A} &=& i\int d^5 x \sqrt{-g} \, (kz) N_A^{(0)} Q \left[ (f_{L+}^{(n)} f_{L+}^{(m)} + f_{R+}^{(n)} f_{R+}^{(m)})
            \overline{\psi}_+^{(n)} \gamma^\mu A_\mu \psi_+^{(m)} + (+\leftrightarrow -)\right],\nonumber\\
     &=& i\int d^4 x \, N_A^{(0)}Q \left(\delta_{mn}-\frac{\eta}{2(kz_{IR})^3}f_{\rm even}^{(m)}(z_{IR})
     f_{\rm even}^{(n)}(z_{IR})\right)\overline{\psi}^{(m)} \gamma^\mu A_\mu \psi^{(n)},
     \label{photonaction}
\end{eqnarray}
where the electric charge $Q =T^3+Y$ and we have used the orthonormal conditions \eqref{e:normgen}.
There are also boundary contributions to the photon couplings which become
\begin{eqnarray}
\label{photonIR}
      \Delta g_{Q-}^{(nm)}&=& \xi_{UV}\, N_A^{(0)}\, Q f_{L-}^{(n)}f_{L-}^{(m)} \bigg|_{z_{UV}} +
    \frac{\xi_{IR}}{(k z_{IR})^3} N_A^{(0)}\, Q f_{L-}^{(n)}f_{L-}^{(m)}\bigg|_{z_{IR}},\\
      \Delta g_{Q+}^{(nm)}&=& \xi_{UV}\, N_A^{(0)}\, Q f_{R+}^{(n)}f_{R+}^{(m)}\bigg|_{z_{UV}}
      + \frac{\xi_{IR}}{(k z_{IR})^3} N_A^{(0)}\, Q f_{R+}^{(n)}f_{R+}^{(m)} \bigg|_{z_{IR}}.
\end{eqnarray}
Therefore, combining the bulk and boundary contributions, the electromagnetic coupling $e$ times the
charge $q$, to the fermion zero modes becomes
\begin{equation}
     e q = N_A^{(0)} Q\left(1-\frac{\eta}{2(kz_{IR})^3}f_{\rm even}^{(0)}(z_{IR})
     f_{\rm even}^{(0)}(z_{IR})\right) + \xi_{UV}\, QN_A^{(0)}\, f_{L-}^{(0)}f_{L-}^{(0)} \bigg|_{z_{UV}} +
    \frac{\xi_{IR}}{(k z_{IR})^3} N_A^{(0)}Q\, f_{L-}^{(0)}f_{L-}^{(0)}\bigg|_{z_{IR}}.
    \label{ecoupling}
\end{equation}
Using the fermion profile solutions (\ref{f0even}) and (\ref{f0odd}), we find the analytic expression
\begin{equation}
      e q=N_A^{(0)}Q\left[1+\frac{\xi_{IR} k - \eta k/2}{1+ (\eta k)/2+(\eta k)^2 \widehat{m}_n^2 }\right]~,
\end{equation}
where the UV boundary contribution is negligible.
Given that $(f_{L-}^{(0)})^2 = (f_{R+}^{(0)})^2$ this coupling is identical for both left and right-handed
fermions. This result is consistent with the requirement that the electromagnetic gauge group remains a
fundamental gauge symmetry in the model. Furthermore, note that only the product (\ref{ecoupling})
of the coupling $e$ times the charge $q$ is determined from (\ref{photonaction}).
Numerically we find that the SM electromagnetic coupling can be reproduced using the values $g_{5L}\sqrt{k} =0.15$,
 $\eta k\sim200$ for light fermions such as electron, $\eta k\sim10$ for the top quark\footnote{The different choice for
 the top quark is related to the fact that $\widehat{m}_t\sim {\cal O}(0.1)$, while for light fermions $\widehat{m}_i\ll1$.} and the gauge sector parameters used for the benchmark point in \cite{cgw}.
 But since $U(1)_Q$ is an exact gauge symmetry in the SM, we need to consider the universality of $e$ and matching to SM charges at high precision level. From (\ref{ecoupling})
we see that with a fixed universal coupling $e$, the slight nonuniversality in the fermion profiles leads to tiny
charge shifts in the usual fermion charges $Q$. In the Standard Model this causes
$U(1)_Q$ violation. For example, since down and strange quarks mix, small
differences in the electric charges of the down and strange quark can ultimately lead to
a photon mass. In addition charge neutrality constraints on the neutron require the
charge shifts to be tiny. A simple way to avoid these constraints is to introduce
flavor dependence into the fermion kinetic coefficients, $\eta$. It is with these flavor dependent coefficients
that the fermion charges take their normal values times a universal electromagnetic coupling. The values
required range from $\delta\eta_e/\eta_e \sim 10^{-7}$ to $\delta\eta_t/\eta_t \sim {\cal O}(1)$.  This tiny
flavor dependence will have negligible effects on other observables, except for the third generation. However
as we will see later, the 3rd generation requires special treatment to be consistent with electroweak precision tests.
Note that for the light fermions the flavor-dependent wavefunction coefficients do lead to flavor-changing neutral
currents (for a nondiagonal 5D Yukawa coupling matrix (\ref{massLag})). These effects are expected to be small and a complete analysis of flavor effects is postponed to future work.
In addition neutrino charge is avoided by tuning the wavefunction coefficients of the gauge bosons so that the
charged electroweak gauge bosons have their usual charge.

\subsection{$W$-boson coupling}

In the emergent model the electroweak gauge symmetry is not fundamental. The $W/Z$ boson masses emerge directly from some sector with strong dynamics and by the holographic duality are identified with the first massive states of a KK tower. In order to be consistent with experimental constraints, it is also necessary to sufficiently separate the first KK mode from the higher KK resonances which can be achieved using boundary kinetic terms. Therefore the expression for the lowest-lying KK mode which
is identified with the $W$-boson is~\cite{cgw}
\begin{equation}
      f_W^{(1)}(z) \simeq \sqrt{\frac{1}{\zeta_L}} \left(\frac{z}{z_{IR}}\right)^2 ,
\label{Wprofile}
\end{equation}
with $m_W= \sqrt{2/(\zeta_Lk)} z_{\rm IR}^{-1}$ and $\zeta_L$ is the coefficient of the $W$-boson boundary kinetic term.
Since the KK mode profiles are not constant, gauge universality is no longer guaranteed and the differences between the fermion wavefunctions lead to non-universality in the gauge couplings.

The couplings to the $W$-boson are obtained by substituting \eqref{KKdec3} into (\ref{gaugeLag}) and
(\ref{boundaryintLag}). They can be separated into couplings to left ($-$) and right-handed ($+$) fermion fields.
The contribution from the bulk is
\begin{equation}
       g_{W\pm}^{(nml)} =  g_{5L} \int_{z_{UV}}^{z_{IR}}\frac{dz}{(kz)^4} f_{iL\pm}^{(n)}f_{jL\pm}^{(m)} f^{(l)}_W,
       \label{Wcoupling}
\end{equation}
while the IR boundary gives the contributions
\begin{equation}
        \Delta g_{W-}^{(nml)}  = g_{5L}\frac{\xi_{IR}}{(k z_{IR})^3}
          f_{iL-}^{(n)}f_{jL-}^{(m)}f_W^{(l)}\bigg|_{z_{IR}}.
        \label{bdyWcoupling}
\end{equation}
Note that the W-boson couplings (\ref{bdyWcoupling}) only receive contributions from the IR boundary
and there are no boundary couplings to right-handed $(+)$ fields, since they vanish at the IR boundary.

We are interested in the 4D gauge coupling of the $W$-boson to the standard model left-handed fermions.
These are identified with the lowest lying KK fermion modes $\psi_-^{(0)}$ with profiles $f_{L-}^{(0)}$. Hence according to the general expressions \eqref{Wcoupling} and \eqref{bdyWcoupling}, the effective 4D gauge coupling for the $W$-boson is
\begin{equation}
      g_W \equiv  g_{W-}^{(001)}+ \Delta g_{W-}^{(001)} =  g_{5L} \int \frac{dz}{(kz)^4}
      f_{iL-}^{(0)}f_{jL-}^{(0)} f_W^{(1)}
      +g_{5L} \frac{\xi_{IR}}{(k z_{IR})^3}   f_{iL-}^{(0)}f_{jL-}^{(0)}f_W^{(1)}\bigg|_{z_{IR}}.
      \label{zeroWcoupling}
\end{equation}
Note also from (\ref{Wcoupling}) that the $W$-boson couples to $\psi_+^{(0)}$. This is an anomalous coupling
to right-handed fermion states. These couplings are generated because the $\Psi^{(L)}$ fields in \eqref{gaugeLag}
are Dirac fermions containing right-handed SU(2)$_L$ doublets. They are a generic prediction of
fermions with boundary mass terms in a warped extra dimension~\cite{neubert}.

To obtain a numerical estimate of the $W$-boson gauge couplings we will consider the bulk fermion example in Sec. 2.2. The gauge coupling (\ref{zeroWcoupling}) becomes
\begin{equation}
g_{W}  \simeq \frac{g_{5L}}{\sqrt{\zeta_L}} \frac{2}{\eta k}
\left[ \frac{1}{3}+ \xi_{IR} k (1 - 2 (\eta k) {\widehat m_i}{\widehat m_j}) \right],
\label{exWcoupling}
\end{equation}
where $i\neq j$. We see that to leading order the gauge coupling is universal, as expected from the universal
fermion bulk profile and boundary fermion kinetic terms. This leading behaviour can be used to determine
the value of the 5D coupling $g_{5L}$. Assuming a fermion kinetic term with coefficient $\eta k = 200$,
$\zeta_L k = 1002$, $\zeta_Y k = 0.2$, $\zeta_Q k = 1715$ and a boundary coupling (\ref{xiIReqn})
with $\Lambda=10 k$, one finds that $g_{5L}\sqrt{k} \sim 0.15$. This value is sufficiently small that the 5D theory remains perturbative.

In addition we see that the nonuniversality in (\ref{exWcoupling}) is quite suppressed,
especially for the lighter fermions where ${\widehat m_i}\ll 1$.
For example, the difference between the tau and muon coupling is one part in $10^4$, while the difference between the muon and electron coupling is negligible. This is consistent with experiment which constrains lepton non-universality at the level of $1/500$ \cite{banj}. Similarly the difference between the up and electron coupling is negligible, consistent with the $1/167$ bound on lepton-quark non-universality from LEP2 \cite{azz}. However the
biggest effects occur for the third generation quarks. The difference between the top-bottom $W$-boson coupling and the up-down $W$-boson coupling is approximately $15\%$, namely
\begin{equation}
\label{Wcouplingratio}
      \frac{g_{W-}(\rm tb)}{g_{W-}(\rm ud)}= 0.854.
\end{equation}
The $Wtb$ coupling has only recently been measured at the $20\%$ level in single top-quark production
at D0~\cite{Abazov:2009ii}.

The anomalous $W$-boson gauge coupling can also be computed analytically. It is found to be
\begin{equation}
        g_{W+}^{(001)}\simeq \frac{g_{5L}}{\sqrt{\zeta_L}} \frac{2}{3}(\eta k)
        {\widehat m_i}{\widehat m_j}\simeq g_W \frac{(\eta k)^2}{3\xi_{IR}k} {\widehat m_i}{\widehat m_j}.
\end{equation}
This coupling is always suppressed compared to $g_W$. The largest value occurs for the third generation
quarks and for the massless bulk fermion example we obtain $g_{W+}^{001}\sim 3\times10^{-4}\, g_W$.

Finally note that we have presented numerical results for massless bulk fermions $(c=0)$. Clearly $c$ can be changed, and in particular we have numerically checked that for $c$ near zero the $W$-boson coupling ratio does not vary greatly.

\subsection{$Z$-boson coupling}
Just like the $W$-boson, the $Z$-boson is a composite state in the emergent model that is identified with the lowest-lying KK state. The expressions for the profile functions
$f_{L3,B}^{(1)}= N_Z {\tilde f}_{L3,B}^{(1)}$
are given by~\cite{cgw}
\begin{eqnarray}
 {\tilde f}_{L3}^{(1)}(z) &=& z\left[J_1(m_Z z) + b_1^{L3} Y_1(m_Z z)\right]
  \simeq\frac{1}{2}m_Zz^2-\frac{m_Z^{-1}}{\zeta_Qk(1+\beta_5^2)},\label{zprofile0}\\
 {\tilde f}_B^{(1)}(z)&=& \frac{N_1^B}{N_1^{L3}} z\left[J_1(m_Z z) + b_1^{B} Y_1(m_Z z)\right]
 \simeq -\frac{\beta_5\log(m_Zz_{\rm IR})}{2\zeta_Qk(1+\beta_5^2)}m_Zz^2-\frac{\beta_5
 m_Z^{-1}} {\zeta_Qk(1+\beta_5^2)},
 \label{zprofile}
\end{eqnarray}
where
\begin{eqnarray}
 b_1^{L3,B}&=&\frac{(\zeta_{L,Y} k)\, {\widehat m}_Z J_1({\widehat m}_Z)-J_0({\widehat m}_Z)}{Y_0({\widehat m}_Z)-(\zeta_{L,Y} k)\, {\widehat m}_ZY_1({\widehat m}_Z)}, \label{coeff}\\
 \frac{N_1^B}{N_1^{L3}}&=&\beta_5\frac{J_1(m_Z z_{UV})+b_1^{L3}\,Y_1(m_Z z_{UV})}{J_1(m_Z z_{UV})+b_1^{B}\,
 Y_1(m_Z z_{UV})},
\end{eqnarray}
with $\beta_5=g_{5L}/g_{5Y}$, ${\widehat m}_Z=m_Z z_{IR}$, $N_Z$ is the normalization factor and
$\zeta_{L,Y}$ are the constant coefficients of the gauge boson IR boundary kinetic terms.

The $Z$-boson coupling to fermions follows from substituting (\ref{KKdec1}) and (\ref{KKdec2})
into (\ref{gaugeLag}) and (\ref{boundaryintLag}). Again the couplings can be split into those involving
left-handed ($-$) and right-handed (+) fermion fields. The contribution from the bulk interactions is
\begin{equation}
       g_{Z\pm}^{(nml)} =   \int_{z_{UV}}^{z_{IR}} \frac{dz}{(kz)^4} \left[\left(g_{5L} f_{L3}^{(l)} T^3
       + g_{5Y} f_{B}^{(l)} Y_{5L}\right)
       f_{L\pm}^{(n)}f_{L\pm}^{(m)} +g_{5Y} f_{B}^{(l)} Y_{5R} f_{R\pm}^{(n)}f_{R\pm}^{(m)}\right],
	\label{Zcoupling}
\end{equation}
while the boundary contributions are
\begin{eqnarray}
      \Delta g_{Z-}^{(nml)}&=& \xi_{UV}\, e_5^2\, Q \left( \frac{f_{L3}^{(l)}}{g_{5L}}
      + \frac{f_B^{(l)}}{g_{5Y}} \right)f_{L-}^{(n)}f_{L-}^{(m)}\bigg|_{z_{UV}}
      + \frac{\xi_{IR}}{(k z_{IR})^3} \left(g_{5L} T^3  f_{L3}^{(l)}
      + g_{5Y} Y_{5L}  f_{B}^{(l)} \right) f_{L-}^{(n)}f_{L-}^{(m)}\bigg|_{z_{IR}},\nonumber\label{anomZm}\\ \\
     \Delta g_{Z+}^{(nml)} &=& \xi_{UV}\, e_5^2\, Q \left( \frac{f_{L3}^{(l)}}{g_{5L}}
     + \frac{f_B^{(l)}}{g_{5Y}} \right)f_{R+}^{(n)}f_{R+}^{(m)}\bigg|_{z_{UV}}
     + \frac{\xi_{IR}}{(k z_{IR})^3}g_{5Y} Y_{5R}  f_{B}^{(l)} f_{R+}^{(n)}f_{R+}^{(m)}\bigg|_{z_{IR}}.
        \label{anomZp}
\end{eqnarray}
We are particularly interested in the $Z$-boson couplings to the lowest-lying fermions.
The couplings $g_{Z-} (g_{Z+})$ to left (right)-handed fermions are therefore given by
\begin{equation}
    g_{Z\mp}\equiv g_{Z\mp}^{(001)} + \Delta g_{Z\mp}^{(001)}.
    \label{ZcouplingLR}
\end{equation}
Note that in addition to the usual $Z$-boson couplings in the standard model, the couplings
(\ref{ZcouplingLR}) of the 5D emergent model contain anomalous couplings. Again this is a
consequence of 5D fermions with brane-localized mass terms. From (\ref{ZcouplingLR}),
(\ref{anomZm}) and (\ref{anomZp}) we see that $g_{Z\pm}$ has anomalous couplings proportional
to $Q$ arising from the UV boundary terms given by
\begin{equation}
    \delta g_{Z\mp}^{(Q)} = \xi_{UV} e_5^2 Q \left.\left(\frac{f_{L3}^{(1)}}{g_{5L}} + \frac{f_B^{(1)}}{g_{5Y}}\right)
    f_{L-,R+}^{(0)}f_{L-,R+}^{(0)}\right|_{UV}.
    \label{UVcoup}
\end{equation}
However on the UV boundary both the gauge boson and fermion profiles are extremely suppressed
so this anomalous coupling is negligible.

The coupling to left-handed fermions, $g_{Z-}$ also contains an anomalous coupling proportional to $Y_R$.
This bulk contribution arises from the left-handed fermions in the Dirac spinor $\Psi^{(R)}$ and is given by
\begin{equation}
     \delta g_{Z-}^{(R)} = g_{5Y} Y_{5R} \int \frac{dz}{(kz)^4} f_B^{(1)} f_{R-}^{(0)}f_{R-}^{(0)}~.
\end{equation}
This contribution is proportional to $\widehat m^2$, so will be suppressed except for the top quark.

Finally the right-handed coupling to fermions, $g_{Z+}$ also has anomalous couplings proportional to $T^3$
and $Y_L$. These are again bulk contributions arising from right-handed fermions in the Dirac spinor $\Psi^{(L)}$. They are given by
\begin{equation}
     \delta g_{Z+}^{(L)} =  \int \frac{dz}{(kz)^4} (g_{5L} T^3 + g_{5Y} Y_{5L})  f_B^{(1)} f_{L+}^{(0)}f_{L+}^{(0)}~.
\end{equation}
Again this contribution is proportional to $\widehat m^2$ and will be suppressed except for the top
quark.

The couplings (\ref{ZcouplingLR}) must be compared with those in the standard model where
\begin{eqnarray}
g_{Z-}^{(SM)}&=&\frac{g}{\cos\theta_W} (\cos^2\theta_W T^3 - \sin^2\theta_W Y_L)~,\\
g_{Z+}^{(SM)}&=&\frac{g}{\cos\theta_W} (-\sin^2\theta_W Y_R)~.
\end{eqnarray}
In terms of the vector $g_V=g_{Z-}+g_{Z+}$ and axial $g_A=g_{Z-}-g_{Z+}$ $Z$-boson couplings to the quarks
one obtains
\begin{equation}
     \frac{g_V^{(SM)}}{g_A^{(SM)}} = 1-4 |Q| \sin\theta_W^2~.
\end{equation}

To match the 5D emergent model to the standard model couplings we use the couplings (\ref{ZcouplingLR}) to obtain the ratio
\begin{equation}
     \frac{g_V}{g_A} = 1+\epsilon-4 |Q| \sin\theta_W^2~,
\end{equation}
where
\beq
\label{epsfn}
      \epsilon = \frac{2\int \frac{dz}{(kz)^4} (g_{5L} {\tilde f}_{L3}^{(1)} - g_{5Y}{\tilde f}_{B}^{(1)})
      f_{L+}^{(0)}f_{L+}^{(0)}}{\int \frac{dz}{(kz)^4} (g_{5L} {\tilde f}_{L3}^{(1)} - g_{5Y}{\tilde f}_{B}^{(1)})
      (f_{L-}^{(0)}f_{L-}^{(0)}-f_{L+}^{(0)}f_{L+}^{(0)})+\frac{\xi_{IR}}{(k z_{IR})^3}(g_{5L} {\tilde f}_{L3}^{(1)} - g_{5Y}{\tilde f}_{B}^{(1)})  f_{L-}^{(0)}f_{L-}^{(0)}\Big|_{z_{IR}}},\\
\eeq
\beq
      \sin^2\theta_W = \frac{-\left(\int \frac{dz}{(kz)^4} g_{5Y}{\tilde f}_{B}^{(1)}(f_{L-}^{(0)}f_{L-}^{(0)}+
      f_{L+}^{(0)}f_{L+}^{(0)})+\frac{\xi_{IR}}{(k z_{IR})^3} g_{5Y}{\tilde f}_{B}^{(1)} f_{L-}^{(0)}f_{L-}^{(0)}\Big|_{z_{IR}}\right)}{\int \frac{dz}{(kz)^4} (g_{5L} {\tilde f}_{L3}^{(1)} - g_{5Y}{\tilde f}_{B}^{(1)})
      (f_{L-}^{(0)}f_{L-}^{(0)}-f_{L+}^{(0)}f_{L+}^{(0)})+\frac{\xi_{IR}}{(k z_{IR})^3}(g_{5L} {\tilde f}_{L3}^{(1)} - g_{5Y}{\tilde f}_{B}^{(1)}) f_{L-}^{(0)}f_{L-}^{(0)}\Big|_{z_{IR}}}.\\
      \label{sintheta}
 \eeq
In deriving the expressions (\ref{epsfn}) and (\ref{sintheta}) we have used the fact that
$(f_{L-}^{(0)})^2 = (f_{R+}^{(0)})^2$. Note also that in the numerator of (\ref{sintheta}) we have neglected the UV boundary couplings (\ref{UVcoup}).

For our example in Section~\ref{c0example} we find the numerical value of (\ref{sintheta}) to be
$\sin^2\theta_W \simeq 0.223$, with negligible deviation amongst the light fermions. This is consistent
with the on-shell scheme value of $\sin^2\theta_W= 1-m_W^2/m_Z^2$. The numerical value
of (\ref{epsfn}) ranges from $\epsilon_e\sim 1.5\times 10^{-13}$ for the electron, up to
$\epsilon_t\sim 9\times 10^{-4}$ for the top quark.

The deviations in the $Z$-boson couplings for the light fermions are consistent with experimental constraints.
For example, the deviation between the tau and electron $Z$-boson coupling is $0.04\%$. The largest deviations in the $Z$-boson couplings occur for the third generation quarks. The deviation in the bottom quark coupling compared to the SM value is at the  $0.3\%$ level.  However for the top quark we obtain
\begin{equation}
\label{Zcouplingratio}
      \frac{g_{Z-}({\rm top})}{g_{Z-}^{(SM)}({\rm top})}= 0.731;  \qquad
      \frac{g_{Z+}({\rm top})}{g_{Z+}^{(SM)}({\rm top})}= 0.732.
\end{equation}
These correspond to deviations in the couplings at the $20\%$ level. This coupling is yet to be
experimentally measured, although at the LHC it will be measured at this level with 300 fb$^{-1}$ of
data~\cite{Berger:2009hi}.

The $Z$-boson coupling deviation can also be converted into a flavor-dependent change of the vector
$g_V=g_{Z-}+g_{Z+}$ and axial $g_A=g_{Z-}-g_{Z+}$, $Z$-boson couplings to the quarks.
Consider first the $b_L$ quark with $(T^3,Q)=(-1/2,-1/3)$.
Numerically we obtain $g_V^{(b)} = -0.264$ and $g_A^{(b)} = -0.376$.
Again using $\widehat m_b = 0.0025$ the corrections are quite small of order one part in  $10^7$.
Furthermore the sign of the corrections does not help to explain the $Zb{\bar b}$ anomaly~\cite{LEP2},
where the experimental values $(g_V^{(b)} ,g_A^{(b)})$ are $( -0.238, -0.38)$ (although this is only for the massless bulk fermion example ($c=0$))
Instead for the top quark $t_L$ with $(T^3,Q)=(1/2,2/3)$ the deviations will again be at the $20\%$ level.

\subsection{Electroweak Precision Analysis}
\label{ewptsec}

In Ref.~\cite{cgw} electroweak precision tests (EWPT) were conducted for the simplified model where all
fermions are confined on the IR brane with a universal gauge coupling. It was found that the non-SM corrections can be compatible with LEP bounds using $z_{IR}^{-1}\sim1.8$ TeV. In the more realistic fermion model
presented here, we have seen that mass-dependent nonuniversality appears in the gauge couplings. Although
the effect from light fermion couplings is negligible, the top-quark couplings gain an ${\cal O}(20\%)$ deviation, which can cause sizeable corrections to EWPT.

As discussed in \cite{Larios:1999au, Berger:2009hi}, the electroweak precision observables most sensitive
to the anomalous top coupling are $\epsilon_1$ (related to the $T$-parameter), and $\epsilon_b$
(related to the $Zb\bar{b}$ coupling). Using the results in \cite{Larios:1999au} (see Eqs. (1), (12) and (13)) we
numerically obtain
\begin{equation}
     \epsilon_1^{SM}+ \delta \epsilon_1 \simeq 19\times 10^{-3}; \qquad
     \epsilon_b^{SM}+\delta\epsilon_b \simeq -13\times 10^{-3},
     \label{epsresults}
\end{equation}
where for a top-quark mass $m_t=171$ GeV we obtain $\epsilon_1^{SM}=9.2\times 10^{-3}$
and  $\epsilon_b^{SM}=-6.1\times 10^{-3}$.
In the SM contribution to $\epsilon_1^{SM}$ we have explicitly subtracted the Higgs contribution
since our model is Higgsless. This compares with the experimental results \cite{altarelli}
\begin{eqnarray}
      4.4\times 10^{-3} \leq &\epsilon_1^{exp}& \leq 6.4\times 10^{-3},\\
      -6.2\times 10^{-3} \leq &\epsilon_b^{exp}& \leq -3.1\times 10^{-3}.
\end{eqnarray}
We see that the results of the model (\ref{epsresults}) are a factor of approximately $2-3$ times outside the current experimental range. However the experimental bounds are at $68\%$ C.L. and increasing to $99\%$ C.L. decreases the discrepancy with our theoretical values.
We have also not included effects from the Kaluza-Klein states, which may compensate our values in (\ref{epsresults}). In addition anomalous top couplings may also generate flavor signals, such as the nonunitarity of the CKM triangle. A detailed discussion of these effects is beyond the scope of this paper but may be just as relevant as the electroweak precision tests.

The origin of the discrepancy with experiment is the fact that the top quark is not as IR localized as the light fermions. As noted in Section 3.1 the deviations are related to the fact that
top quark has a large mass close to IR scale: ${\widehat m}_t\sim 0.1$. A simple way to avoid the large deviations for the top quark
is to increase the IR scale $z_{IR}^{-1}$ up to $\sim 10$ TeV.
This leads to ${\widehat m}_t\sim{\cal O}(0.01)$, causing the IR boundary contribution to dominate. The non-unversality between the top quark and other fermion couplings can then be reduced to be within experimental bounds. However such a high IR scale would typically not unitarize WW scattering at the TeV scale.
Therefore the only way to reasonably account for the discrepancy with experiment is to treat the third generation quarks differently compared to the light fermions. In fact similar issues have also been encountered with other Higgsless models such as technicolor or the 5D Higgsless models \cite{topcolor, Csaki:2003zu}. Indeed in these models the problem is addressed by separating the physics of electroweak symmetry breaking from the physics which generates the top quark mass, where the scale of the top quark mass generation is at a higher scale to reduce the anomalous top coupling~\cite{topcolor,Csaki:2005vy,Cacciapaglia:2005pa}. Therefore in our framework, one possibility would be to have two separate bulk AdS spaces meeting at the IR brane (TeV scale), one for the top quark and the other for the light fermions. The AdS bulk for the top quark has a smaller compactification radius, so that the IR scale for that bulk is ${\cal O}(10)$ TeV. The details of this plausible solution are left for future work.

\section{Conclusion}

We have shown how fermion masses are generated in a model of electroweak symmetry
breaking with composite $W,Z$ gauge bosons. While the generation of $W,Z$ boson masses
does not rely on a Higgs mechanism, resulting instead from the breaking of conformal symmetry,
the fermions are mostly composite states that emerge at the IR scale and obtain a mass from electroweak symmetry breaking at the Planck scale. Gauge coupling universality is achieved via a universal fermion
profile and the suppression of both gauge boson and fermion profiles at the UV brane.
The fermion mass hierarchy is no longer generated by the wavefunction overlap of bulk profiles,
but instead relies on a Froggatt-Nielsen mechanism near the UV scale.

In summary, all light fermions fit well in our current model--the generation of mass, universal gauge
couplings and agreement with electroweak precision tests. Although there is no conflict with
current direct bounds, the $20\%$ deviation of the top-quark coupling from the SM leads to a tension
with indirect bounds from electroweak precision data. Nevertheless these problems can be averted
by treating the top quark separately like in other Higgsless models. Of course there may be other
possibilities where all fermions can be treated in a uniform framework, and this is open to future
model building. Furthermore, the lightest Kaluza-Klein top quark in our setup has a mass of 1.5 TeV
and this state can lead to an interesting discovery signal at the LHC. The idea that the electroweak gauge
symmetry is not fundamental and mass is generated without a Higgs mechanism remains an intriguing
possibility. It will soon be put to further stringent tests at the LHC.

\section*{Acknowledgements}
We thank Zhenyu Han, Alex Pomarol and James Wells for helpful discussions.
Y.C. is supported by NSF grant PHY-0855591 and the Harvard Center for Fundamental Laws of Nature.
T.G. and J.S. are supported by the Australian Research Council.

\end{document}